\documentclass[aps,prd,twocolumn,notitlepage,nofootinbib,nobibnotes,superscriptaddress,amsmath,amssymb,preprintnumbers]{revtex4-1}

\usepackage{graphicx} 
\usepackage{dcolumn}
\usepackage{bm}        
\usepackage{amssymb}   
\usepackage{adjustbox}  
\usepackage{amsmath,array}
\usepackage{comment}
\usepackage{natbib}
\usepackage{MnSymbol}
\usepackage[hidelinks]{hyperref}
\usepackage{graphicx}
\usepackage[caption=false]{subfig}
\usepackage{color}
\usepackage     [utf8]                  {inputenc}
\usepackage     [T1]                    {fontenc}
\usepackage     [english]               {babel}
\usepackage{epigraph}
\usepackage{etoolbox}
\usepackage{ dsfont }
\usepackage{physics}
\usepackage{xcolor}
\usepackage{siunitx}
\makeatletter
\patchcmd{\epigraph}{\@epitext{#1}}{\itshape\@epitext{#1}}{}{}  
\newcommand*\eqsize{%
\@setfontsize\mysize{9.0}{9.0}%
    }
    \usepackage{multirow}
\makeatother

\usepackage{xfrac}
\usepackage{amsmath,amssymb}
\usepackage{esdiff} 
\usepackage{commath} 
\usepackage{bbm} 
\usepackage{braket}
\usepackage{slashed}

\newcommand{\rmd}{\mathrm{d}}

\newcommand{\rmi}{\mathrm{i}}
\newcommand{\rmq}{\mathrm{q}}

\newcommand{\fm}{\mathrm{fm}}

\newcommand{\qqbar}{\rmq\bar{\rmq}}

\newcommand{\xp}{\boldsymbol{x}_\perp} 			
\newcommand{\pp}{\boldsymbol{p}}
\newcommand{\qp}{\boldsymbol{q}}

\newcommand{\GeV}{\;\text{GeV}}
\newcommand{\MeV}{\;\text{MeV}}
\newcommand{\TeV}{\;\text{TeV}}

\definecolor{oscar}{RGB}{22, 156, 172}
\definecolor{peter}{RGB}{255,0,0}

\begin{document}

\date{\today}

\title{Anomalous dilepton production from the Dirac sea in p+p collisions}

\author{Oscar Garcia-Montero}
\email{garcia@fias.uni-frankfurt.de}
\affiliation{Institut f\"{u}r Theoretische Physik,
                     Goethe Universit\"{a}t, Max-von-Laue-Strasse 1,
                     60438 Frankfurt am Main, Germany}

\begin{abstract} 
In a recent work from the ALICE collaboration an excess of relatively soft dilepton pairs was reported in p+p collisions at $13\TeV$. In the report no satisfactory explanation was found via standard bremsstrahlung computations. In this letter, \textit{anomalous} dilepton pairs are produced non-perturbatively from the back-reaction of the vacuum to a $\qqbar$ pair, travelling back-to-back. The dilepton spectrum is computed by folding this rate with the cross-section for the process p+p$\rightarrow\qqbar$, which was computed using the framework of the Color Glass Condensate Effective Field Theory (CGC EFT) in the $k_\perp$-factorized limit. The resulting spectra is compatible within uncertainties with the data, and moreover, provides exciting insight into the non-perturbative phenomena that ultra-soft electromagnetic radiation can probe. 
\end{abstract}

\maketitle

\section{Introduction}

In the setting of Heavy Ion Collisions, electromagnetic radiation is an excellent probe of the space-time evolution. Because of the lack of strong interactions, and final-state effects, it can escape unscathed from the interaction volume. It is then important to emphasize that because of the same reasons, photons and dielectrons are good messengers of information of strong non-perturbative phenomena at smaller momentum scales. Effects such as chiral dynamics, which are not yet fully understood then provide a promising avenue of research.

This may be the case for the consistent excess of electromagnetic radiation, also called  \textit{anomalous radiation}, measured across different hadronic collision experiments \cite{[{See, for example: }]Chliapnikov:1984ed, *Botterweck:1991wf,*Banerjee:1992ut,*Belogianni:2002ic,*Belogianni:1997rh}, as well as by the DELPHI collaboration at the Large Electron Positron Collider \cite{Abdallah:2005wn,*Abdallah:2010tk}. Standard perturbative photon production was not enough to explain the then novel signal. Adding to this are the recent findings of the ALICE collaboration in p+p collisions at $\sqrt{s}=13\TeV$, where a softer part of the phase space is accessible thanks data taking in a lower magnetic field \cite{Acharya:2020gjz,Acharya:2020rfm,*Acharya:2018kkj,*Acharya:2018ohw}. 
Apart from the bremsstrahlung channels, in medium hadronic production \cite{Rapp:2013nxa} was included in the comparison, but the total yield obtained by this model was not enough to account for the excess.  

Given the very soft degrees of freedom accessed by the ALICE collaboration, a possible explanation of the phenomenon would be that the leptonic pairs are produced by quarks of gluon origin. This is supported by the fact that gluons should dominate the hadronic wave function in the kinematic range accessed at midrapidity. Electromagnetic bremsstrahlung from quarks was computed by adapting the results of Ref. \cite{Benic:2018hvb,Benic:2016uku} and proved to be insufficient to account for the excess. This hints that the pairs are produced by non-perturbative means, meaning that the pairs are not irradiated from a quasi-particle final state like a hadron or an on-shell parton. Instead, one could look at the model proposed by Kharzeev and Loshaj \cite{Kharzeev:2013wra}, where radiation is produced by the the back-reaction of the vacuum to a quark and antiquark pair,  back-to-back. The traveling pair triggers meson-like excitations, which couple directly to photons. Based on the fair description of the DELPHI photon excess in that work, I would like to propose a source for soft $e^+e^-$ production based on this idea. However, instead of producing the $\qqbar$ pairs via decay of neutral $Z_0$ bosons, in the present work they are produced by gluon fusion. This is performed in the framework of the Color Glass Condensate Effective Field Theory (CGC EFT), which is an effective theory of Quantum Chromodynamics (QCD) in the high-energy limit, valid for small-$x$ degrees of freedom. The CGC captures the rise in the gluon abundance, and saturation rises naturally, described by the appearance of a dynamically generated semi-hard scale $Q_s$ \cite{Gelis:2010nm}. Using the CGC framework allows us to create the $\qqbar$ pairs via non-collinear gluon distributions, which has an impact in the spectra of the pair \cite{Fujii:2006ab,Blaizot:2004wu,Blaizot:2004wv}, and therefore on the final shape of the dilepton spectrum. 

This work is organized as follows. First, I give a small review of the underlying model based on Ref. \cite{Kharzeev:2013wra}, followed by a brief description of quark-antiquark production in the CGC EFT. A numerical results are discussed for the comparison with ALICE data for p+p collisions at 13TeV. Finally, I give a brief summary and outlook. This letter will be followed by a more extense and theoretically inclined paper \cite{LongPaper2021}.

\section{Anomalous dilepton production}
In this setting, the electromagnetic degrees of freedom  are radiated from the back-reaction of the vacuum, excited by a $\qqbar$ pair, travelling back-to-back. The current, produced non-perturbatively, is the operator that couples to the virtual photon, producing dileptons via the expression
\begin{equation}
\frac{\rmd N_{e^+e^-}}{\rmd M^2\rmd^2 k_\perp\rmd Y  } = \frac{1}{2\,(2\pi)^3} \frac{\kappa(M)}{M^2} \left|\, j_{tot}^\mu\,j_{tot,\mu}^*\right|^2\,,
\label{eq:dieleptons0}
\end{equation}
where the phase-space function $\kappa(M)$ is 
\begin{equation}
\kappa(M)=\frac{\alpha_{e}}{3\pi}\left( 1+\frac{2m_e^2}{M^2}\right)\sqrt{1-\frac{4m_e^2}{M^2}}\,.
\end{equation}

The dilepton momentum is fully characterized using the invariant mass, $M$,  transverse momentum, $\boldsymbol{k}_\perp$, and rapidity $Y$. Since the computation of this current is extremely complex in full QCD glory, a simpler 1+1D abelian model, is used, namely the massless QED$_{2}$. This model, also called the Thirring model, is given by the Lagrangian function,
\begin{equation}
\mathcal{L} = -\frac{1}{4}\, G_{\mu\nu}\, G^{\mu\nu} + \bar{\psi}\rmi(\slashed{\partial}+ig\slashed{B})\psi-g\, j_{ext}^\mu B_{\mu}\,,
\label{eq:QED2}
\end{equation}
where $\mu=0,1$ and $B_\mu$ is the abelian gauge field which is a proxy for the gluons.
It is important to note that even though this model is abelian, it shares some properties with QCD, i.e. it is confining. In eq. \eqref{eq:QED2}, the external current, 
\begin{equation}
j^\mu_{ext}(x)=e\,q_f\,[ u_+^\mu\,\delta (z-v t)\,\theta(z) - u_-^\mu\,\delta (z+v t)\,\theta(-z)]\,,
\end{equation}
corresponds to massless quark and antiquark, traveling with velocity vectors $u_\pm^\mu=(1,\pm v)$. Here, the speed of the (anti) quark is $v=p/\sqrt{Q_0^2+p^2}$, where $p$ is their momentum, and $Q^0\sim 2$ GeV is a timelike virtuality scale. The latter acts an effective mass to the leading quark pair, as it is roughly the threshold at which the pQCD partonic stops and confinement starts. In what follows, I make the association $x^0 = t$ and $x^1 = z$ for simplicity.  

It is widely known that such a model can be solved by the process of bosonization \cite{Coleman:1975pw}, in which the system is simplified by introducing a scalar field and making the associations, 
\begin{equation}
\begin{split}
j^\mu(x)=\bar{\psi}(x) \gamma^\mu \psi(x)  = -\epsilon^{\mu\nu}\partial_\nu \phi(x)/\sqrt{\pi}\\
j^{5\mu}(x)=\bar{\psi}(x) \gamma^\mu \psi(x)  = -\partial_\nu \phi(x)/\sqrt{\pi}
\end{split}
\label{eq:currents}
\end{equation}
for which one then obtains a free, but externally driven, scalar theory. For the massive case, bosonization yields a more complex Sine-Gordon scalar \cite{Mandelstam:1975hb,Coleman:1974bu}. The equation of motion for the relevant degrees of freedom is then simply
\begin{equation}
(\partial^2+m^2)\,\phi(x) = -m^2 \,\phi_{ext}(x)
\label{eq:KG}
\end{equation}
where the scalar mass is defined to be $m^2=g^2/\pi$. The external source, $\phi_{ext}$, can be found by inverting eq. \eqref{eq:currents} for $j_{ext}$. To get a simple analytical form, the Klein-Gordon equation can be solved in the ultra-relativistic limit, $v\rightarrow 1$, which solved in Milne coordinates, $\tau=\sqrt{t^2-z^2}$ and $y=\log[(t+z)/(t-z)]/2$, becomes independent of $y$. The  solution fort the resulting equation of motion is 
\begin{equation}
\phi(m\,\tau) = -\sqrt{\pi}\,\theta(m^2\,\tau^2)\,(1- J_0(m\,\tau))\,,
\end{equation}   
from where the corresponding charge density can be extracted, 
\begin{equation}
j^0(\tau,y) = -m\,\theta(m^2\,\tau^2)\,\sinh y\,J_1(m\,\tau)\,.
\end{equation}  

Here $J_i$ stands for the $i$th Bessel function of the first kind. This oscillation of charge density is also associated (via eq. \eqref{eq:currents}) to oscillation of axial charge. It is this this non-perturbatively created current that couples to the electromagnetic $U(1)$ and produces (virtual) photons. Finally, one can express the total current, $j_{tot}^\mu=j_{ext}^\mu + j^\mu$, in momentum space, as it is given in Ref. \cite{Kharzeev:2013wra}
\begin{equation}
j_{tot}^\mu(k) =- \rmi e q_f \frac{2\,v\,\epsilon^{\mu\nu}k_\nu}{k^2_0-v^2k^2_3}\, \left( 1+ \frac{m^2}{k^\mu k_\mu-m^2}\right)
\end{equation}
where $e$  is the electromagnetic coupling with $\alpha=e^2/4\pi$. Here, we set the Levi-Civita to $\epsilon^{01}\equiv 1$. The current is comprised of two parts, the left term in the parenthesis comes from standard bremsstrahlung of the leading current, and the right term, which is the additional back-reaction of the vacuum. 
To couple this expression to electromagnetic radiation in 3+1D, one needs to associate the momentum modes of the current, $k$, with the momentum of the outgoing virtual photon. This can be done by absorbing the virtual photon's $\boldsymbol{k}_\perp$,  into $M$, via $k^\mu k_\mu=k_0^2-k^2_z \rightarrow M^2 + k_\perp^2\equiv M_\perp$, where $M_\perp$ is the transverse mass of the pair. The production rate in terms of the pair variables is given by 
\begin{equation}
\begin{split}
\frac{\rmd N_{e^+e^-}}{\rmd M^2\rmd^2 k_\perp\rmd Y  } = \frac{\alpha}{(2\pi)^2}&\frac{\kappa(M)}{M^2}\,\frac{4\,v^2\,M^2_\perp}{(k^2_0-v^2k^2_3)^2}\\
\times &\left( 1+ \frac{m^2}{M^2_\perp-m^2}\right) ^2\,.
\end{split}
\label{eq:fullee}
\end{equation}

If one takes the limit $M\rightarrow 0$, one recovers the photon properties  such as the fact that the rate vanishes with $k_\perp$. This is quite important, as it means that the Low theorem is satisfied \cite{Low:1958sn}, which requires that as $k_\perp\rightarrow 0$, only asymptotic states can radiate photons. 

There are still two more assumptions one has to take before this rate can be phenomenologically used, which I will briefly explain, but the interested reader can find a more extensive explanation in Ref. \cite{Kharzeev:2013wra,LongPaper2021}. First, since the meson-like state $\phi$ is interpreted to be an effective overlap of different hadronic states, i.e. contains the hadronic spectrum, the mass cannot be fixed. In the Thirring model, a pair od static fermions are pulled together at small distances by a linear potential $\sim \pi m^2/2=\kappa_s$, where $\kappa_s$ is the string tension. In this work gaussian fluctuations are assumed for the string tension parameter, where $\langle\kappa_s^2\rangle = 0.9\GeV/\fm$, as it is suggested in the literature \cite{Bali:1992ab, *Pirner:2018ccp}. 

The second consideration rises as the simplified model does not include the decay of the mesons, so it has to be included by upgrading the denominator so that $m^2 \rightarrow m^2 -\rmi \gamma^2$. This new parameter is related to the decay rate via $\gamma^2 = m\Gamma$ and it
has been fitted to a value of $3.7^{+0.8}_{-0.7} \MeV $ using the photon excess from the DELPHI data \cite{Abdallah:2005wn}. To do so, one has to take the $M\rightarrow 0$ limit, and compute the total photon number per $\qqbar$ pair. This result is consistent with Ref. \cite{Kharzeev:2013wra}, as with the Particle Data Group, which locates $\gamma$ for the neutral isoscalar particles between $8\times10^{-4}\GeV$ and $8\times10^{-2}\GeV$ \cite{Beringer:1900zz}.

\subsection*{$Q\bar{Q}$ pair prodution in the CGC framework}

If one wants to compute the total radiation produced in a p+p collision, $\qqbar$ pairs have to be sampled from a distribution that accounts for its creation. In this work I have chosen to create the quark-antiquark pairs from gluoproduction, because of the enhancement of gluon distributions at low-$x$ degrees of freedom. To do so, I use the $k_\perp$-factorized cross-section found in Refs. \cite{Gelis:2003vh,Blaizot:2004wv}, which was computed in the CGC EFT formalism. I use here the $k_\perp$-factorized form because our system, $p+p$, can be considered a dilute-dilute system. A dilute hadron is such that the characteristic momentum of the gluon distribution is larger than the saturation scale of the hadron,$k_\perp^2 > Q_s^2$. The cross-section used in this limit is given by
\begin{equation}
\begin{split}
\frac{\rmd \sigma^f_{\qqbar}}{\rmd^2\qp\,\rmd y_q\,\rmd^2\pp\,\rmd y_p}  = &\frac{\alpha_S^2}{16\pi^4 \,C_F}\int \frac{\rmd^2\boldsymbol{k}_{1,\perp}}{(2\pi)^2} \frac{\varphi_1(x_1,\boldsymbol{k}_{1,\perp})}{\boldsymbol{k}^2_{1,\perp}}\\ &\times\frac{\varphi_2(x_2,\boldsymbol{k}_{2,\perp})}{\boldsymbol{k}^2_{2,\perp}}\,\Theta(\boldsymbol{k}_{1,\perp} ,\boldsymbol{k}_{2,\perp})
\end{split}
\label{eq:CGC}
\end{equation}
where $\alpha_S$ is the strong coupling constant, and $k_{i\perp}, \varphi_i$ are the gluon transverse momentum and unintegrated distribution function (uGDF) for the $i$th proton, respectively. The kinematic variables used here are $\boldsymbol{k}_{2,\perp}=\pp+\qp -\boldsymbol{k}_{1,\perp}$, $x_1 \sqrt{s/2} =  p^- + q^- $ and $x_2 \sqrt{s/2} = p^+ + q^+$ where the light-cone variables are defined as $p^\pm=\sqrt{\pp_\perp^2+m_q^2} e^{\pm y_p}/\sqrt{2}$. The uGDFs are given explicitly by 
\begin{equation}
\frac{\varphi_i(x_i,k_{i,\perp})}{k_{i,\perp}^2} = \frac{S_\perp  }{2\alpha_S} \int \rmd^2 \xp\,e^{\rmi \xp \cdot k_{i,\perp}} \left\langle U(\xp)U^\dag(0) \right\rangle\,,
\label{eq:phi}
\end{equation}
where $S_\perp$ represents the transverse area of the proton. The uGDFs are expressed in terms of the adjoint Wilson lines $U(\xp)$, and the $x$ dependence is contained in the weight function used in the averaging. Finally, $\Theta(\boldsymbol{k}_{1,\perp} ,\boldsymbol{k}_{2,\perp})$ represents the fermionic hard factors, which can be found in Refs. \cite{Gelis:2003vh,Blaizot:2004wv}. The total dilepton spectrum is  obtained by folding the $\qqbar$ cross-section, eq. \eqref{eq:CGC}, with the single-pair production rate, eq. \eqref{eq:dieleptons0}. However, there is a small caveat to take into account.
While the dilepton rate was computed in a purely longitudinal, center-of-mass (CM) frame of the $\qqbar$ pair, the distribution function samples (anti) quark momenta in the laboratory frame. The two can be reconciled by first boosting the lab-frame to the CM frame 
\begin{equation}
k_{cm}^\mu =  {\Lambda_z}^\mu_\nu(-\beta_z) \,{\Lambda_x}^\nu_\lambda(-\beta_x)\,k^\lambda
\end{equation}
where $\beta_z =\tanh Y$. The transverse boost is performed along the $x$ direction without loss of generality, using $\beta_x=k_\perp/\sqrt{M^2 +k_\perp^2}$ (see Ref. \cite{Fujii:2006ab}). In the CM frame, however, the quark's momentum doesn't need to be purely longitudinal, as it can be parametrized as 
\begin{equation}
q^\mu_{cm}= \left(M/2, q_{cm}\, \cos(\varphi),q_{cm}\, \sin(\varphi), \sqrt{M^2/4-q_{cm}^2-m_e^2}\right)\,. 
\end{equation}
where $q_{cm}$ and $\varphi$, are the CM transverse momentum and azimuthal angle. For the antiquark an extra minus sign is added in the spatial components. Using this form, one has to further rotate, as to align the virtual photon to the longitudinal direction used in the model, 
\begin{equation}
k_{l} = R^{-1}_{xz}(q_{cm,\perp},q_{cm,z})\,R^{-1}_{xy}(\varphi_{cm})\,k_{cm}\,,
\end{equation}
where $k_{l}$ is the input momentum used to sample from eq. \eqref{eq:fullee}. Finally, the longitudinal momentum of the (anti) quark is then equated to $p$ in the definition of the leading $\qqbar$ pair velocity $v$.

\section{Results}

The following results were computed numerically using VEGAS, a Monte Carlo integration algorithm with importance sampling. The $x_i$ dependence of the adjoint dipoles, eq. \eqref{eq:phi}, is computed using the running coupling Balitsky-Kovchegov (BK) equation \cite{Dusling:2009ni}, which gives a good approximation to the JIMWLK hierarchy \cite{Balitsky:1995ub,Kovchegov:1999yj}. The initialization for such dipoles is given at    $x_0=0.01$ by the McLerran-Venugopalan model with an anomalous dimension of  $\gamma=1.13$ and an initial saturation scale of $Q^2_0=0.168\GeV^2$. It is worth noting that these parameters give good description of data according with the fit of ref. \cite{Albacete:2012xq}. To account for the high-$x$  region of the cross-section, the uGDFs are matched to the collinear gluon parton distribution function (PDF) at $x=x_0$ using the prescription of Ref. \cite{Ma:2014mri}. From this method the proton transverse radius is extracted, $R_p=0.48\fm$, which is consistent with the literature \cite{Ma:2018bax}. This is used to compute the transverse area $S_\perp$. The quark masses were set to  $m_u = m_d = 0.005\GeV$, $m_s = 0.095\GeV$. For consistency, I only take on account the $u,d,s$ system, as the underlying model assumes massless quarks. The reason for his is that characteristic scale of the uGDF, and therefore the gluon-fusion process in the cross-section is of order $\langle k^2_{\perp}\rangle \gtrsim Q^2_0$, and only the quarks listed above have masses safely below this characteristic scale.

\begin{figure}
\centering
   \includegraphics[scale=0.44]{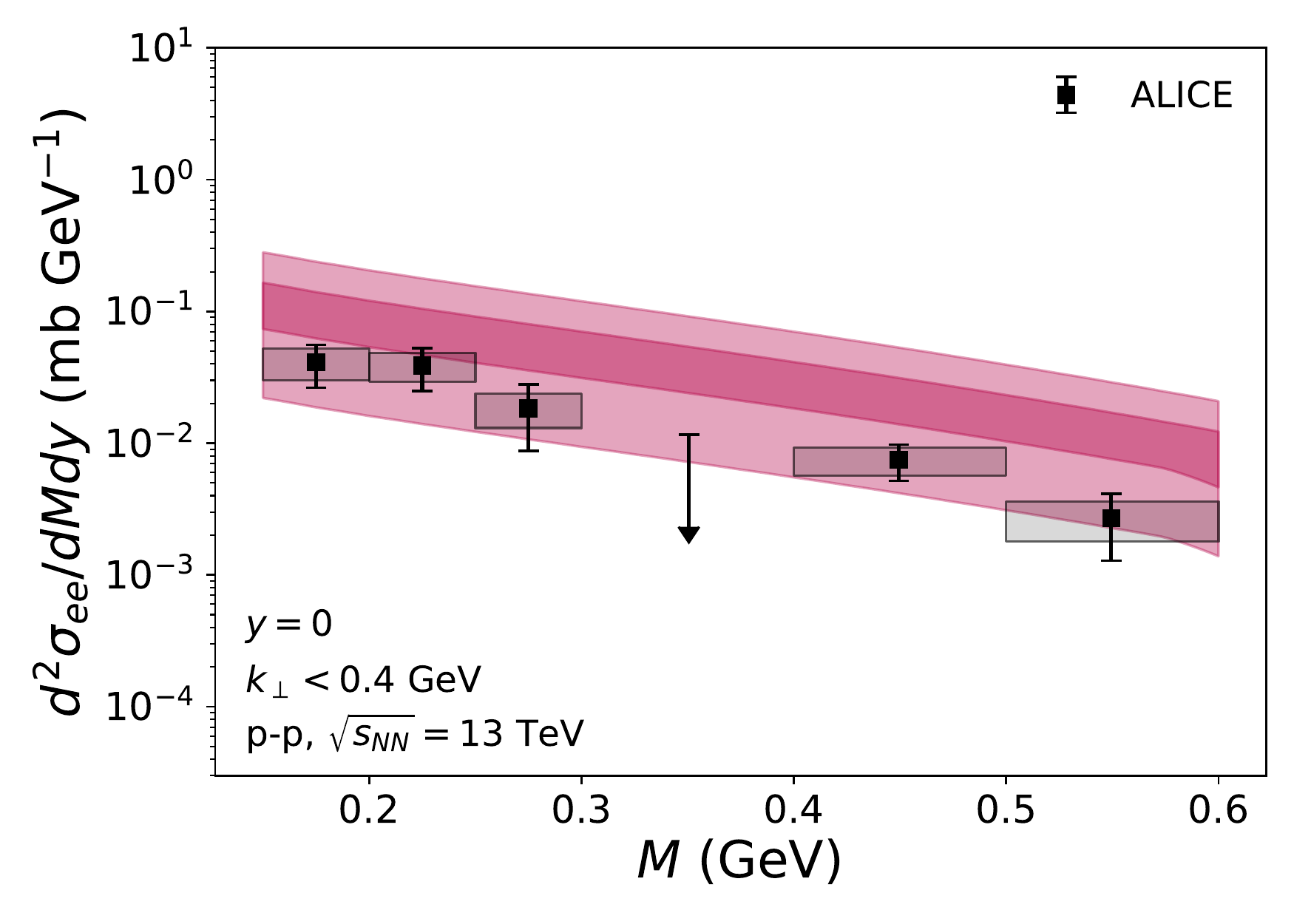}
   \includegraphics[scale=0.45]{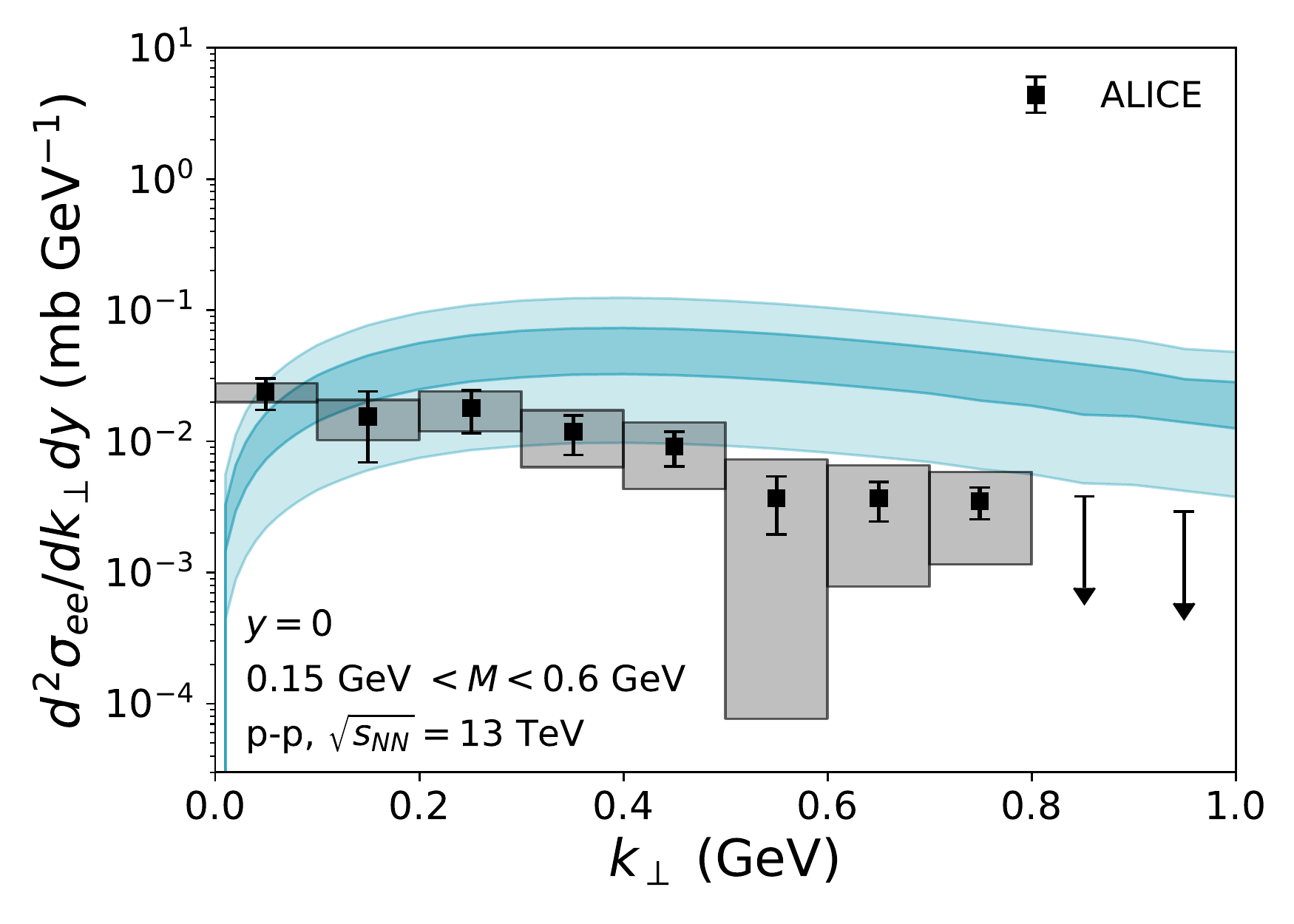}
    \caption{Comparison of the total anomalous spectrum to the dilepton excess reported in Ref.  \cite{Acharya:2020gjz} at midrapidity, for different kinematic ranges. \textit{Up:} Invariant mass spectrum, integrated over $k_\perp < 0.4\GeV$. \textit{Down:} pair momentum spectrum, integrated over the range $0.15\GeV < M < 0.6\GeV$. The darker bands correspond to variations on $\gamma$ allowed by the fitting procedure, while the lighter bands are associated to systematic uncertainties in the CGC computations. }
\label{fig:data}
\end{figure}

The yield was computed for the kinematic region $0.15\GeV <M<0.6$ GeV, $k_\perp<1\GeV$ and $|Y|<0.8$ and was further integrated along $M,k_\perp$ for the respective cuts in Fig. \ref{fig:data}. To account for the uncertainties of the calculation two different bands were computed. The darker bands comprise the bounds of $\gamma$ compatible to the fit to DELPHI data. The lighter band accounts for the uncertainties in the $\qqbar$ cross-section, namely in the BK truncation of the JIMWLK hierarchy, whose error is of order $N_c^{-1}$, and the determination of $S_\perp$. The latter comes from the fit of the CGC to the  collinearly factorized PDFs, and because of the quadratic dependence of the cross-section to the transverse area, the error is known to amount up to 50\% \cite{Benic:2018hvb,Ma:2018bax}. The lighter band then amounts for an extra $60\%$ uncertainty.

 The resulting spectra were compared to the excess signal from Ref. \cite{Acharya:2020gjz}. The reader can observe in Fig. \ref{fig:data} that this simple model achieves a fair agreement with data. It is important to note that this model should be taken as a qualitative study, as it lacks the quark mass, which limits the computation to the lighter quarks. This problem can be solved by upgrading the model to massive QED$_2$, which can also be solved by bosonization. The resulting Sine-Gordon model is not analytically solvable, but it may be systematically computed by mass-perturbation theory \cite{Adam:1997wt}, or numerically. Furthermore, including the spacetime rapidity in eq. \eqref{eq:KG} is fundamental to better understand the shape of the  dilepton spectra.
 
\section{Summary and Conclusions}

In this work I present the computation of soft dilepton pairs radiated from non-perturbative dynamics of $\qqbar$ pairs, produced by gluon fusion processes. The production of these dileptons relies on the oscillation of charge which arises from the leading quarks ripping meson-like fermion states from the Dirac sea.  I have computed the spectrum for the kinematics of the dilepton excess published by the ALICE collaboration, and found fair agreement between this simplified model and the data.

Because of the simplifications taken in this model there are shortcomings which affect the results. Nevertheless, this the model can be systematically improved in the future by including both rapidity dependence in the pair CM, as well as non-vanishing mass quarks. Furthermore, modelling this system with real-time dynamical simulations may help to better understand systems closer to QCD, as well to explore the dynamics of photoproduction via string-breaking \cite{Spitz:2018eps}, as well as during a turbulent thermalization of the surrounding medium \cite{Berges:2014bba, *Berges:2015ixa,*Berges:2013fga}. 

As final remark, given the consistent finding of anomalous radiation across different experiments, it seems important to delve into more measurements of soft electromagnetic radiation. This will give rise to the possibility to start to understand the truly non-perturbative dynamics surrounding the strong  interaction. 

\section{Acknowledgements}
I would like to thank Jürgen Berges, Daniel Spitz and Yunxin Ye, as well as Hannah Elfner and Horst Sebastian Scheid for very enlightening discussions. I would like to thank Jerome Jung for talking me into this topic. This project was supported by  the Deutsche Forschungsgemeinschaft (DFG, German Research Foundation) – Project number 315477589 – TRR 211. Computational resources have been provided by the GreenCube at GSI.
\bibliographystyle{apsrev4-1}
\bibliography{AnomalousDilepton.bib}

\end{document}